\documentclass[12pt]{iopart}

\usepackage{graphicx}% Include figure files

\begin{document}

\title[]{Dependence of the nature of the Holstein polaron motion in a polynucleotide chain subjected to a constant electric field on the initial polaron state and the parameters of the chain}

\author{A.N. Korshunova$^1$ \& V.D. Lakhno$^2$}

\address{Institute of Mathematical Problems of Biology, Russian Academy of
Sciences, \\ Pushchino, Moscow Region, 142290, Russia}
\ead{$^1$alya@impb.ru, $^2$lak@impb.ru}
\vspace{10pt}
\begin{indented}
\item[]September 2021
\end{indented}

\begin{abstract}
In this work, we consider the motion of a polaron in a polynucleotide Holstein molecular chain in a constant electric field. It is shown that the character of the polaron motion in the chain depends not only on the chosen parameters of the chain, but also on the initial distribution of the charge along the chain. It is shown that for a small set value of the electric field intensity and for fixed values of the chain parameters, changing only the initial distribution of the charge in the chain, it is possible to observe either a uniform movement of the charge along the chain, or an oscillatory mode of charge movement.
\end{abstract}


%\maketitle

%\maketitle

\section {Introduction}

A large number of theoretical and experimental works are devoted to the study of the charges movement in molecular polynucleotide chains \cite{Zhongkai} -- \cite{9}.
The urgency of this problem is associated with the possibility of using one-dimensional polynucleotide chains as nanowires in nanobioelectronic devices \cite{10} -- \cite{14}. It is believed that the main current carriers in such chains are self-trapped electronic states, which have the form of polaron formations \cite{15} -- \cite{Lak2000}.

In this study, the features of charge movement in homogeneous molecular polynucleotide chains are considered. The simulation of the polaron motion was carried out in the presence of a constant electric field in the chain based on the Holstein model \cite{Holstein,Holstein2}. Despite the simplicity of the chosen model, various and complex dynamic modes can be realized in the system under consideration, since this problem is a multiparametric problem. The nature of the charge movement along the chain depends on many parameters of the system: on the parameters of the chain, on the magnitude of the electric field intensity, on the initial distribution of the charge in the chain.

Previous studies (\cite{jtf1} -- \cite{LaKo2007}) show that the uniform motion of a polaron along a chain is possible for small values of the electric field intensity.  The possibility of uniform motion of a charge in a homogeneous Holstein chain in a constant electric field over very long distances was shown in the work \cite{jtf1}. At large values of the electric field intensity, uniform motion is not observed, the charge loses its initial shape and moves along the chain in the direction of the field in the mode of oscillatory motion with Bloch oscillations.

In this paper, it is shown that the nature of the movement of the polaron along the chain depends not only on the value of the electric field intensity, but also on the set parameters of the chain and on the initial charge distribution. To simulate the movement of a charge in a homogeneous polynucleotide chain, such a value of the electric field intensity was chosen at which there is a uniform motion of a polaron in the chain.

It is shown that at a fixed value of the electric field intensity and at fixed values of the chain parameters, changing only the initial charge distribution, it is possible to observe completely different types of charge distribution along the chain. That is, depending on the initial charge distribution, either a uniform charge movement or an oscillatory mode of motion can be observed. It is also shown that, depending on the selected parameters of the chain, during the Bloch oscillations, the polaron can retain its shape for some time. For other parameters, the initial polaron can quickly disintegrate, and then the charge moves along the chain in the direction of the field, performing Bloch oscillations.

\section{Mathematical model}
The dynamic behavior of a polaron in the presence of a constant external field in a homogeneous molecular chain is modeled by a system of coupled quantum-classical dynamic equations with dissipation. In our model, DNA is considered as a homogeneous chain composed of $N$ sites. Each site is a nucleotide pair, which is considered as a harmonic oscillator \cite{Lak2000}. To simulate the dynamics of a quantum particle in a chain of $N$ nucleotide pairs, we will use the Holstein Hamiltonian,	where each site is a diatomic molecule \cite{Holstein,Holstein2}:

The dynamics of a quantum particle in a classical chain is described by a system of nonlinear differential equations, which in dimensionless variables has the form:
\begin{eqnarray}
i\frac{db_n}{d\tilde{t}}&=&-\eta\bigl(b_{n+1}+b_{n-1}\bigr)+\kappa\omega^2u_nb_n+Enb_n, \label{5}\\
\!\!\!\frac{d^2u_n}{d\tilde{t}^2}\!&=&\!-\omega'\frac{du_n}{d\tilde{t}}-\omega^2u_n-|b_n|^2\,,\label{6}
\end{eqnarray}
where $b_n$ -- are the amplitudes of the probability of charge's occurrence at the $n$-th site, $\sum_n |b_n(\tilde{t})|^2=1$, $\eta$ -- are matrix elements of the transition over the sites,  $\omega$ -- is the frequency of oscillations of the  $n$-th site,  $kappa$ -- is the coupling constant, $\omega'$ -- is a friction coefficient,  $u_n$ -- are displacements of sites from their equilibrium positions,   $E$ -- is the electric field intensity,   $\tilde{t}=t/\tau$, $\tau=10^{-14}$ sec (arbitrary time scale).

Equations \eref{5} are Schr\"{o}dinger equations for the probability amplitudes  $b_n$, which describe the evolution of a particle in a deformed  chain. Equations \eref{6} represent classical motion equations which describe the dynamics of nucleotide pairs with allowance for dissipation.
The system of nonlinear differential equations  \eref{5}), \eref{6} is solved by fourth-order Runge -- Kutta method.  The calculations were performed using the computing facilities of the Interdepartmental Center for Supercomputing,
Russian Academy of Sciences.

In this work, to simulate the movement of a charge in an electric field, the following values of the dimensionless parameters were chosen: the coupling constant $\kappa=4$, the matrix elements of the transition over the sites $\eta=2.4$.

The stationary solution of the equations \eref {5}, \eref {6} in the absence of an external electric field corresponds to a function in the form of an inverse hyperbolic cosine:
\begin{equation}\label{7}
|\,b_n(0)|=\frac{\sqrt{2}}{4}\sqrt{\frac{\kappa}{|\,\eta|}}\,\mathrm{ch}^{-1}
\Bigl(\frac{\kappa(n-n_0)}{4|\,\eta|}\Bigr).
\end{equation}
For the chosen values of the parameters $\kappa=4$ and $\eta=2.4$, the initial polaron state of the form \eref{7} slightly differs from the steady polaron for the given chain.
We call a polaron steady, which does not shift from its position in the chain and does not change its shape in the absence of an electric field or additional excitations in the chain.

We will set the initial values of the function $|\,b_n(0)|$ in the form of a stretched inverse hyperbolic cosine:
\begin{eqnarray}\label{7_1}
|\,b_n(0)|=\frac{\sqrt{2}}{4}\sqrt{\frac{\kappa}{\xi|\,\eta|}}\,\mathrm{ch}^{-1}
\Bigl(\frac{\kappa(n-n_0)}{4\xi|\,\eta|}\Bigr),\\
u_n(0)=|\,b_n(0)|^2\big/\omega^2,\ \  du_n(0)\big/d\tilde{t}=0. \nonumber
\end{eqnarray}
where $\xi$ -- is the stretching coefficient, with the help of which we can choose the initial polaron of the form of \eref{7_1} as close as possible to the steady polaron, and also, we can take the initial polaron narrower or wider than the steady one for the formation of various variations of the charge motion along the chain. Thus, an expression of the form of \eref{7_1}, with a correctly chosen value of $\xi$, can be considered an approximate solution to the stationary solution of equations \eref{5}, \eref{6}. For a chain with parameters  $\kappa=4$ and $\eta=2.4$ the inverse hyperbolic cosine or initial polaron of the form of \eref{7_1} is as close as possible to the steady polaron for $\xi=0.95$.

\begin{figure}[h]
\center
\resizebox{0.4\textwidth}{!}{\includegraphics{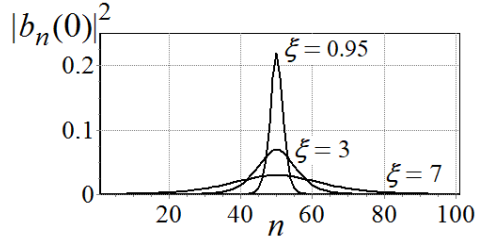}}
\caption{Graphs of the functions $|b_n(0)|^2$ of the form of \eref{7_1} for $\xi=0.95$, $\xi=3$ and $\xi=7$. The values of the chain parameters $\kappa=4$, $\eta=2.4$.}\label{Bn0}
\end{figure}

Figure \ref{Bn0} shows graphs of the functions $|b_n(0)|^2$ of the form of \eref{7_1} for $\xi=0.95$,  $\xi=3$ and $\xi=7$. The graph of the function $|b_n(0)|^2$ at $\xi=0.95$ practically coincides with the corresponding function of the steady polaron in the given chain, so we can say that in Figure \ref{Bn0} for $\xi=0.95$ the graphs of the probability function for a steady polaron are shown.

To simulate the motion of a polaron in a constant electric field, we will place in the chain an initial polaron state of the form \eref{7_1} at the required values of the stretching coefficient $\xi$. We place the center of the polaron on the site of the chain with the number  $n_0$. The value of  $n_0$  is chosen so that at the beginning of the calculations the polaron be far enough from the ends of the chain. Similarly, the length of the chain is chosen so that at the end of the calculations the polaron would not come too close to the end of the chain. The field turns on "instantly" at the initial moment of time.

\section{Dependence of the nature of the polaron motion on the initial polaron state and the parameters of the chain}

In all the examples presented below, we used chains with fixed values of the coupling constant $\kappa$ and matrix elements of the transition over the sites $\eta$. The parameters $\kappa$ and $\eta$ determine the shape of the steady polaron in the chain, which is close to the inverse hyperbolic cosine of the form \eref{7}. Also, in all examples, the same value of the electric field intensity $E$ was used. The value of the electric field intensity and the value of the parameters $\kappa$ and $\eta$ determine the characteristics of the Bloch oscillations, namely: the period of the Bloch oscillations $T_{BL}=2\pi/E$ and the maximum Bloch amplitude $A_{BL}=4\eta/E$. Thus, in all the examples presented below, we have the same characteristics of Bloch oscillations and the same shape of the steady polaron.

Such chain parameters as the oscillation frequency of the $n$th site $\omega$ and the coefficient of friction $\omega'$ were changed. Various initial polaron states of the form \eref{7_1} were also used for $\xi=0.95$, $\xi=3$ and $\xi=7$. The following examples are shown for large values of the parameters $\omega=1$, $\omega'=1$ and for small values $\omega=0.01$, $\omega'=0.006$.

To simulate the motion of a charge in an electric field, the following values of dimensionless parameters were selected: $\kappa=4$, $\eta=2.4$. The dimensionless value of the electric field intensity was set by $E=0.016$. With this value of $E=0.016$, there is a possibility of uniform motion of the polaron in the chain. The value of the electric field intensity and the value of the parameter $\eta$ determine the characteristics of the Bloch oscillations, namely: the period of the Bloch oscillations $T_{BL}=2\pi/E$ and the maximum Bloch amplitude $A_{BL}=4\eta/E$. For $E=0.016$, the period of Bloch oscillations is $T_{BL}=2\pi/E\approx393$, the maximum Bloch amplitude is $A_{BL}=4\eta/E\approx600$.

\begin{figure}[h]
\center
\resizebox{0.37\textwidth}{!}{\includegraphics{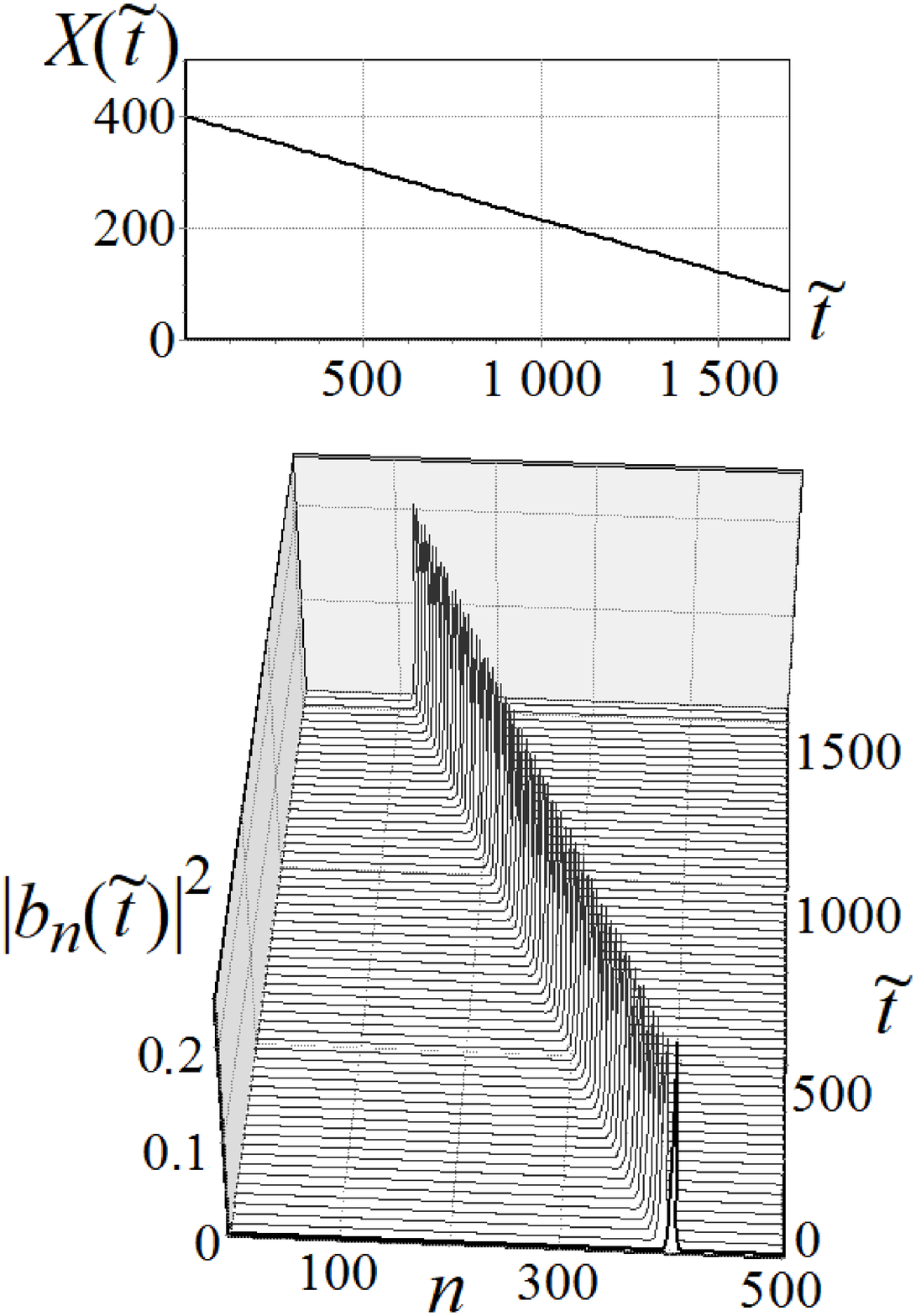}}\textbf{(a)}
\resizebox{0.37\textwidth}{!}{\includegraphics{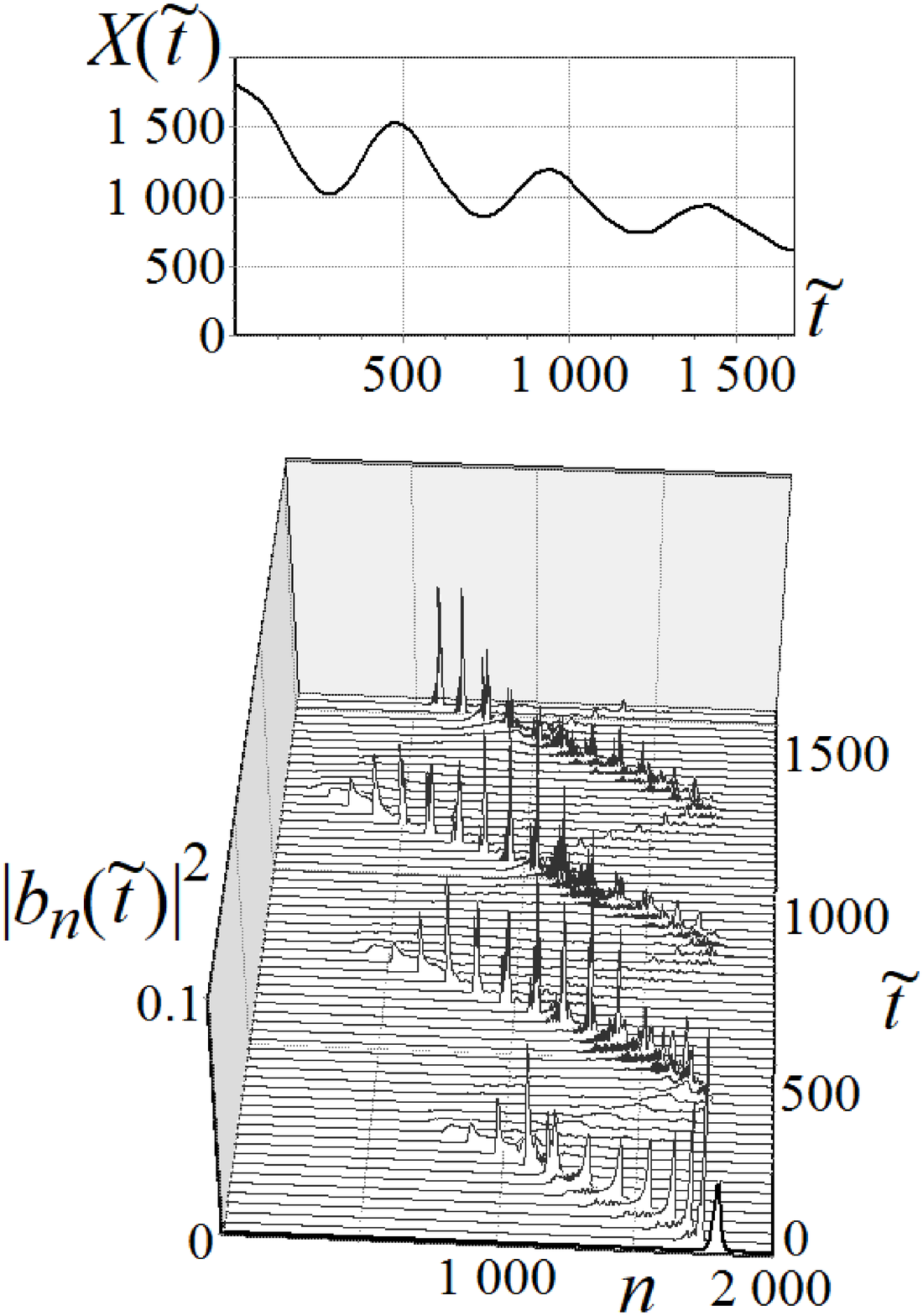}}\textbf{(b)}
\caption{Graphs of the functions $X(\widetilde{t})$ and $|b_n(\widetilde{t})|^2$  during the motion of a polaron in a chain with parameters $\kappa=4, \eta=2.4$, $\omega=1, \omega'=1$. The electric field intensity is $E=0.016$. In Figure (a), the initial polaron state of the form \eref{7_1} is set at $\xi=0.95$, in Figure (b) - at $\xi=7$.}\label{ww11_0.95_7}
\end{figure}

The Figure \ref{ww11_0.95_7} shows graphs of the functions $|b_n(\widetilde{t})|^2$ and graphs of the functions $X(\widetilde{t})$ describing the motion of the center of mass of the particle:
\begin{equation}\label{8}
X(\widetilde{t})=\sum\nolimits_{n}{|\,b_n(\widetilde{t})|^2}\cdot n.
\end{equation}

In the examples in \Fref{ww11_0.95_7} the following values of the chain parameters are selected: $\kappa=4, \eta=2.4, \omega=1, \omega'=1$. In the \Fref{ww11_0.95_7}(a), the center of the initial polaron state is located at the site with the number $n_0=100$ in a chain of length $N=501$. In the \Fref{ww11_0.95_7}(b) -- on a site with the number $n_0=1800$ in a chain of length $N=2001$. The length of the chain in these examples does not significantly affect the nature of the charge distribution along the chain, the length is set differently for a more visual representation of the graphs of functions. Thus, presented in Fig. \ref{ww11_0.95_7} examples, differ only in the form of the initial polaron state. In the \Fref{ww11_0.95_7}(a), the initial values $|\,b_n(0)|$ were chosen in the form of an inverse hyperbolic cosine of the form \eref{7_1} for $\xi=0.95$, in \Fref{ww11_0.95_7}(b) -- for $\xi=7$.

The initial polaron in \Fref{ww11_0.95_7}(a) is as close as possible to the steady polaron in the chain, which is shown in \Fref{Bn0} for $\xi=0.95$. The initial polaron in \Fref{ww11_0.95_7}(b) is set to be stretched, see \Fref{Bn0} for $\xi=7$.

In the example in \Fref{ww11_0.95_7}(a), the initial polaron is set as close as possible to the steady polaron in the chain. Presented in \Fref{ww11_0.95_7}(a) graphs of the functions $X(\widetilde{t})$ and $|b_n(\widetilde{t})|^2$ show that the polaron, retaining its shape, moves uniformly in the electric field. The function $X(\widetilde{t})$ looks like a straight line, which indicates a uniform movement.

If the initial polaron state is set to be stretched, see \Fref{ww11_0.95_7}(b), then we are no longer observing a uniform motion of the polaron, but an oscillatory mode of charge motion. In this case, the charge moves in the direction of the field and performs oscillations, the characteristics of which are slightly different from the Bloch ones, namely: the oscillation period and the maximum amplitude are slightly larger than the theoretical values of the corresponding characteristics of the Bloch oscillations, for a given value of the electric field intensity.

\begin{figure}[h]
\center
\resizebox{0.37\textwidth}{!}{\includegraphics{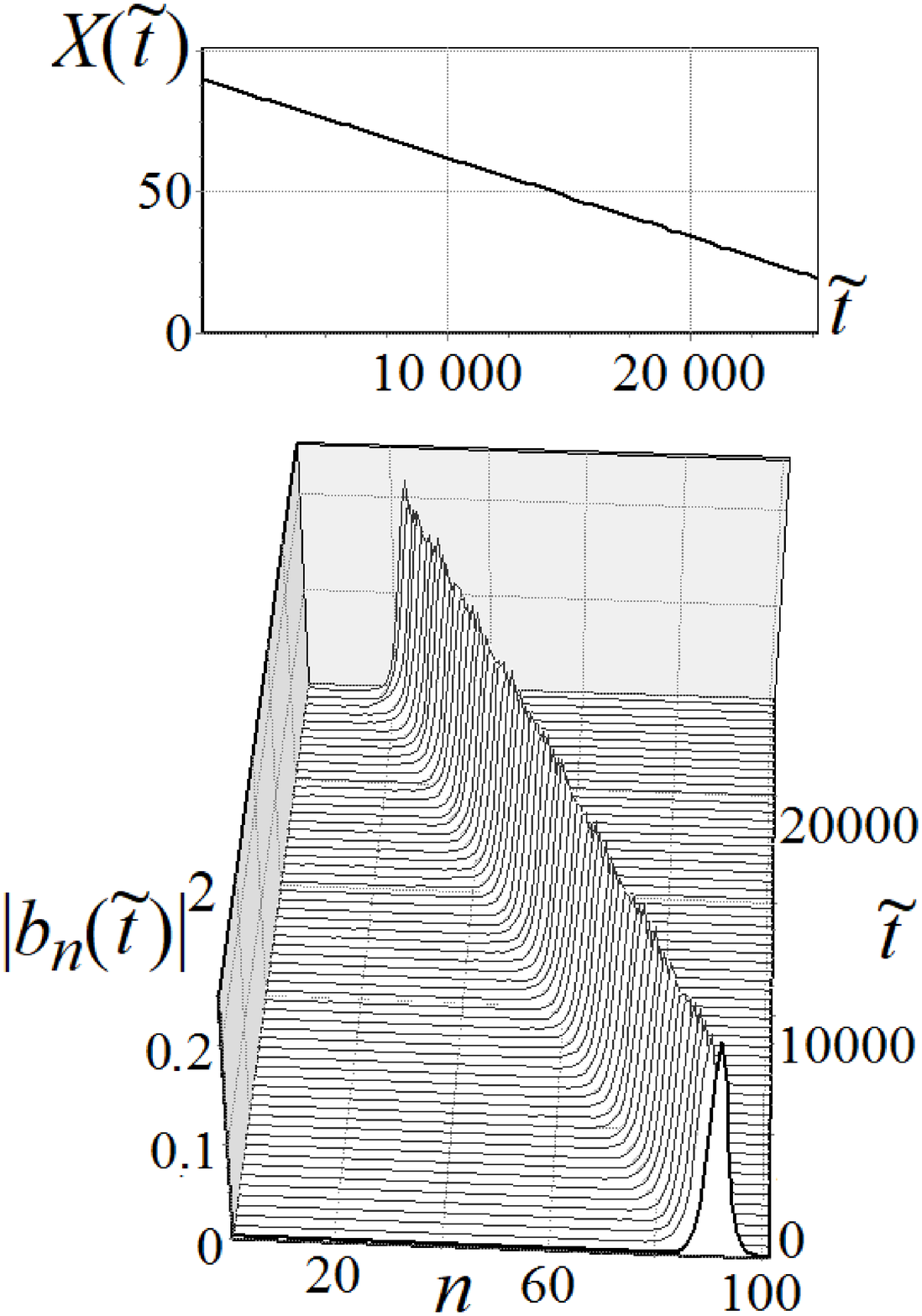}}\textbf{(a)}
\resizebox{0.38\textwidth}{!}{\includegraphics{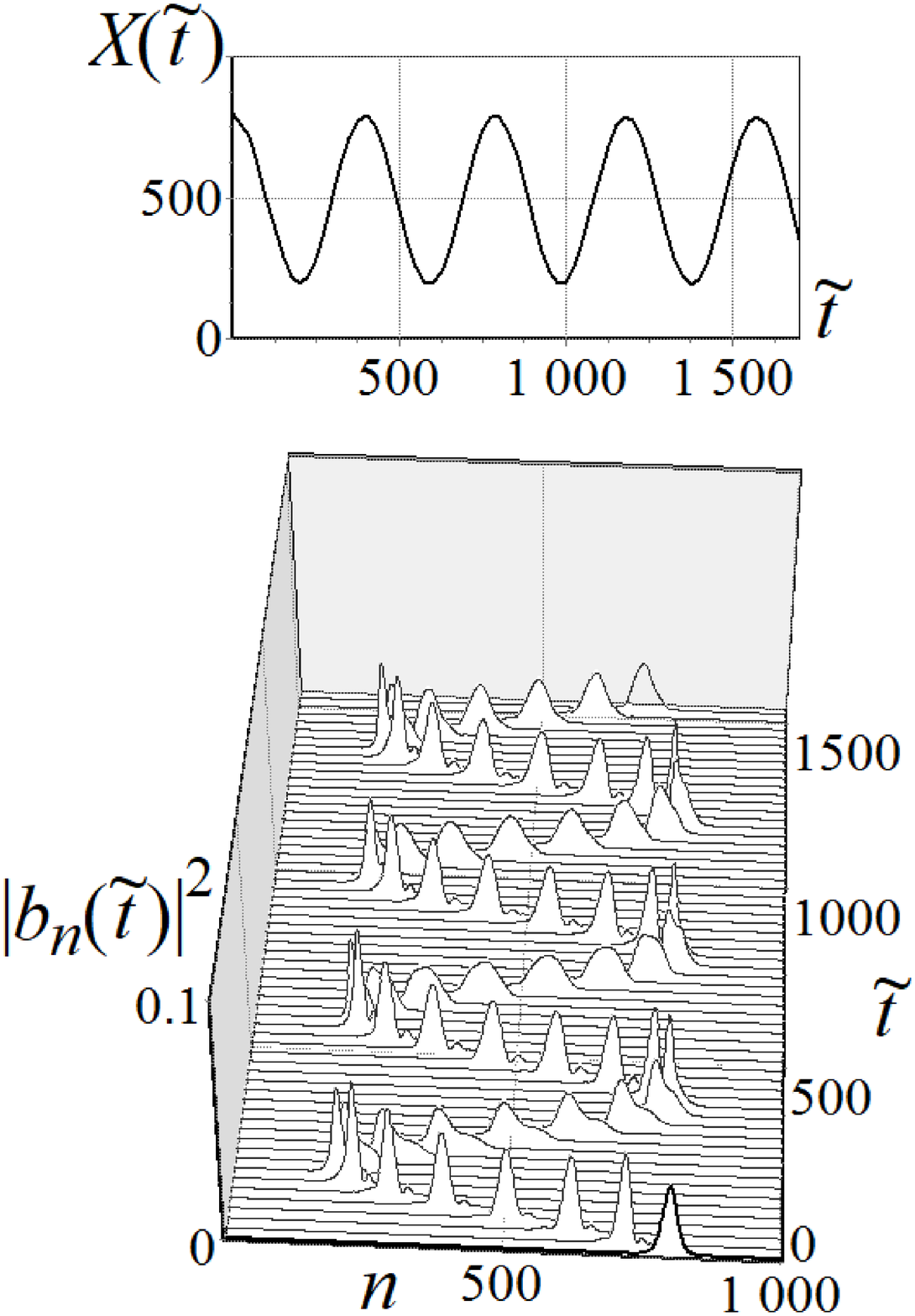}}\textbf{(b)}
\caption{Graphs of the functions $X(\widetilde{t})$ and $|b_n(\widetilde{t})|^2$  during the motion of a polaron in a chain with parameters $\kappa=4, \eta=2.4$, $\omega=0.01, \omega'=0.006$. The electric field intensity is $E=0.016$. In Figure (a), the initial polaron state of the form \eref{7_1} is set at $\xi=0.95$, in Figure (b) - at $\xi=7$.}\label{ww_dna_0.95_7}
\end{figure}

Examples in \Fref{ww_dna_0.95_7} differ from the examples in \Fref{ww11_0.95_7} only with the values of the parameters $\omega$ and $\omega'$. In the examples in \Fref{ww_dna_0.95_7} small values are set: $\omega=0.01, \omega'=0.006$. The initial polaron state in \Fref{ww_dna_0.95_7}(a) is the same as in \Fref{ww11_0.95_7}(a), that is, the initial values of $|\,b_n(0)|$ are given in the form of an inverse hyperbolic cosine of the form \eref{7_1} for $\xi=0.95$. With such small values of the parameters ($\omega=0.01, \omega'=0.006$), the initial polaron also moves uniformly along the chain, keeping its shape. But the velocity of the polaron in this case is tens times less.

In \Fref{ww_dna_0.95_7}(b) and \Fref{ww11_0.95_7}(b), the initial polaron state is set stretched, in the form of an inverse hyperbolic cosine of the form \eref{7_1} for $\xi=7$. Presented in \Fref{ww_dna_0.95_7}(b) graphs of the functions $X(\widetilde{t})$ and $|b_n(\widetilde{t})|^2$ indicate an oscillatory mode of motion with Bloch oscillations. But, unlike the example in \Fref{ww11_0.95_7}(b), the initial polaron state does not fall apart immediately, the charge moves along the chain in the direction of the field, approximately retaining its shape for a sufficiently long time. In this case, for small values of the parameters $\omega=0.01, \omega'=0.006$, the oscillation period and the maximum amplitude of the charge oscillations are very close to the theoretical values of the corresponding characteristics of the Bloch oscillations, for a given value of the electric field intensity, see \Fref{ww_dna_0.95_7}(b).

\begin{figure}[h]
\center
\resizebox{0.37\textwidth}{!}{\includegraphics{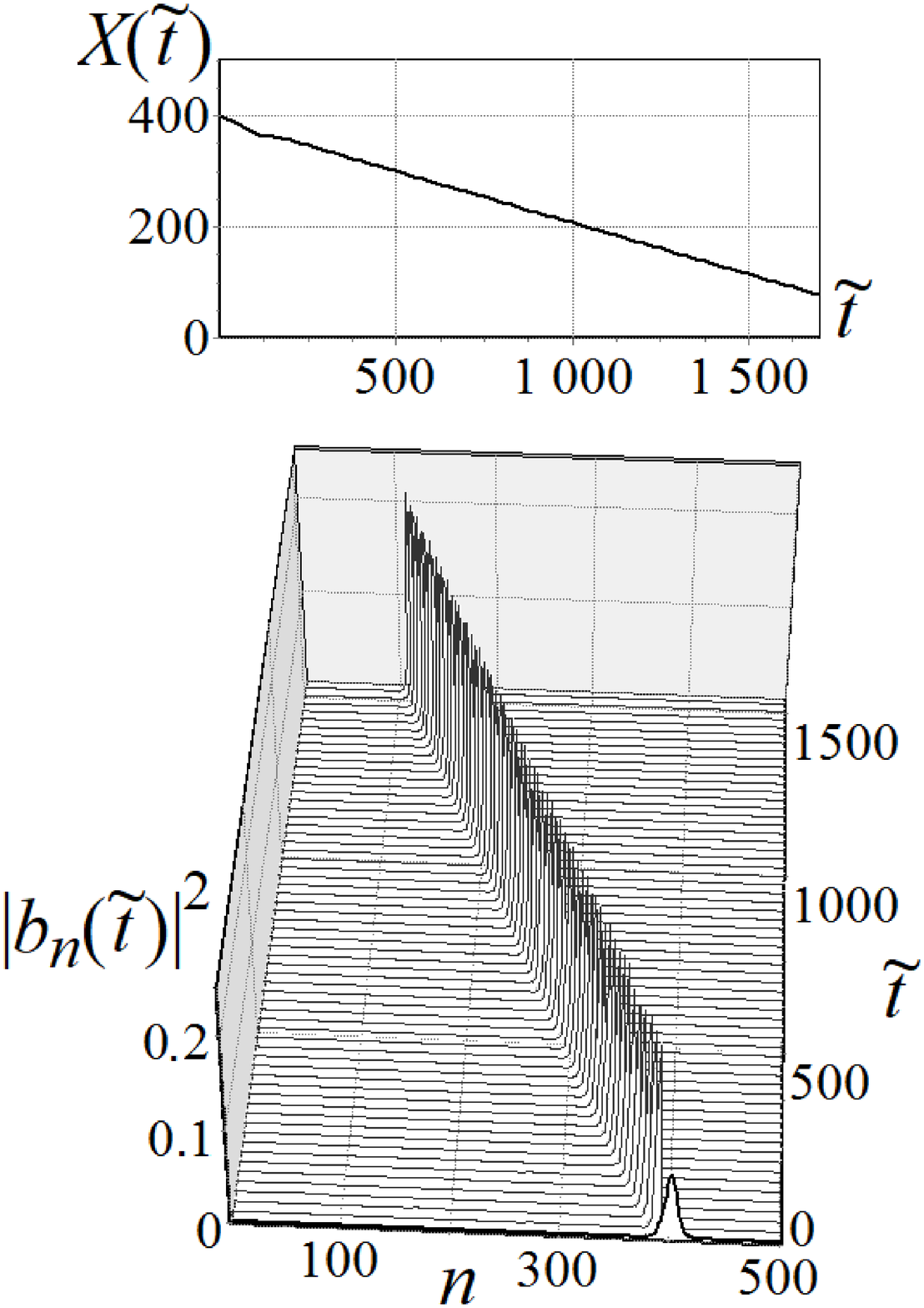}}\textbf{(a)}
\resizebox{0.37\textwidth}{!}{\includegraphics{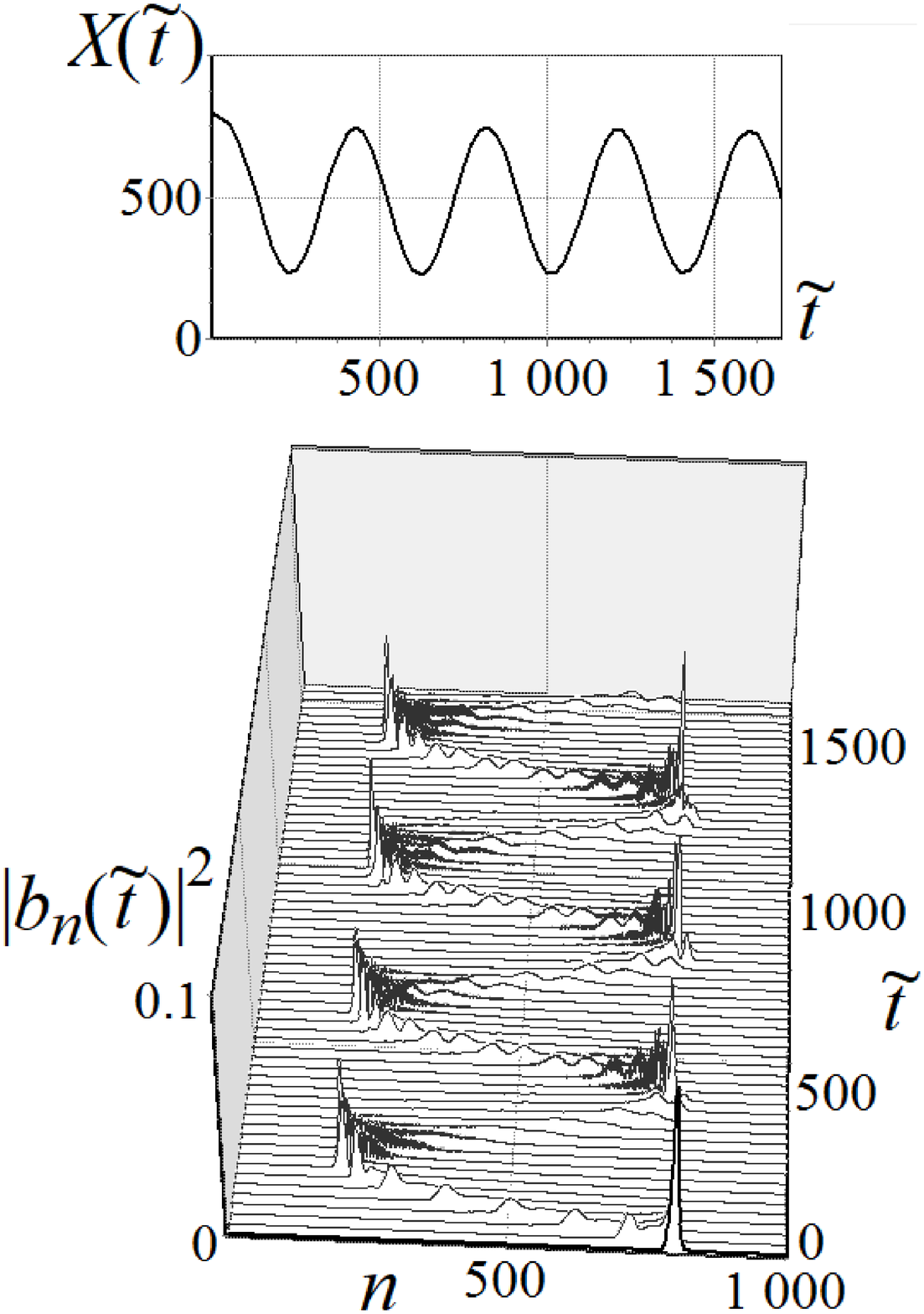}}\textbf{(b)}
\caption{Graphs of the functions $X(\widetilde{t})$ and $|b_n(\widetilde{t})|^2$  during the motion of a polaron in a chain with parameters $\kappa=4, \eta=2.4$ and the initial polaron state of the form \eref{7_1} for $\xi=3$. The electric field intensity is $E=0.016$. In Figure (a), the parameter values are $\omega=1, \ \omega'=1$, in Figure (b) -- $\omega=0.01,\ \omega'=0.006$.}\label{ww_11_dna_3}
\end{figure}

In the examples in \Fref{ww_11_dna_3} the same values of the parameters of the chains $\kappa=4, \eta=2.4$ and the same value of the electric field intensity $E=0.016$ are used as in the previous examples. In the examples in \Fref{ww_11_dna_3}(a) and \Fref{ww_11_dna_3}(b) the same initial values $|\,b_n(0)|$ of the form \eref{7_1} are set for $\xi=3$, see \Fref{Bn0} for $\xi=3$. Presented in \Fref{ww_11_dna_3} examples, differ only in the values of the parameters of the chains $\omega$ (oscillation frequency of the $n$-th site) and $\omega'$ (coefficient of friction).

For large values of the parameters $\omega=1$ and $\omega'=1$ in \Fref{ww_11_dna_3}(a) the stretched initial polaron state of the form \eref{7_1} at $\xi=3$ quickly takes a form close to the form of an steady polaron.  Then the polaron moves along the chain at a constant velocity, retaining its shape. The further motion of the polaron in this case is no different from the motion of the polaron in the example in \Fref{ww11_0.95_7}(a).

In the example in \Fref{ww_11_dna_3}(b) small values of the oscillation frequency of the $n$-th site and the friction coefficient are used: $\omega=0.01, \omega'=0.006$. Graphs of the functions $X(\widetilde{t})$ and $|b_n(\widetilde{t})|^2$ in \Fref{ww_11_dna_3}(b) indicate an oscillatory mode of motion with Bloch oscillations. In this case, as in the example in \Fref{ww_dna_0.95_7}(b), the period of oscillations and the maximum amplitude of charge oscillations are very close to the theoretical values of the similar characteristics of Bloch oscillations, for a given value of the electric field intensity $E=0.016$: period $T_{BL}=2\pi/E \approx 393$ and the maximum Bloch amplitude $A_{BL}=4\eta/E \approx 600$.

Examples in \Fref{ww_11_dna_3}(b) and in \Fref{ww_dna_0.95_7}(b) differ only in the shape of the initial polaron state and demonstrate an oscillatory mode of charge motion. These two examples clearly show that only a change in the shape of the initial polaron state causes a significant change in the nature of the motion and the distribution of the charge along the chain. Graphs of functions $|b_n(\widetilde{t})|^2$ in \Fref{ww_11_dna_3}(b) show that a narrower initial polaron state of the form \eref{7_1} at $\xi=3$ quickly loses its original shape. For a wider initial polaron state of the form \eref{7_1} at $\xi=7$ in the example in \Fref{ww_dna_0.95_7}(b), the charge, oscillating, moves along a chain in the direction of the field, approximately retaining its shape in the initial period of time.

\section{Conclusions}

Previous studies show that in a homogeneous polynucleotide chain there is a possibility of uniform motion of a charge in a constant electric field over very long distances, see \cite{jtf1}. With an increase in the value of the electric field intensity, uniform motion is not observed, the charge passes into a mode of oscillatory motion with Bloch oscillations.

In this work, the simulation of charge motion in a chain was carried out for a single value of a constant electric field: $E=0.016$. With this value of the electric field intensity, the charge can move along the chain at a constant velocity. In addition to the fixed value of the electric field intensity, we used fixed values of the
coupling constant $\kappa=4$ and matrix elements of the transition over the sites $\eta=2.4$. The specified chain parameters ($\kappa$ and $\eta$) determine the shape of the steady polaron in the chain and the main characteristics of the Bloch oscillations for a given value of the electric field intensity.

It is shown that the velocity of the polaron motion in the case of uniform charge motion and the nature of the charge distribution along the chain in the oscillatory regime of motion strongly depend on the set values of the oscillation frequency of the sites $\omega$ and the coefficient of friction $\omega'$.

It is shown that in a chain with fixed values of all parameters, at the same value of the electric field strength, changing only the initial distribution of the charge, it is possible to observe either a uniform motion of the charge or an oscillatory mode of motion.

Thus, the calculations performed show that complex dynamic regimes, which strongly depend on the set of all parameters of the system, can be realized in the system under consideration. By changing the value of at least one of the system parameters, it is possible to modify the charge motion and charge distribution along the chain.

\ack {
This study was performed using the computational
resources of the Interdepartmental Center for Supercomputing,
Russian Academy of Sciences.\\
The work was done with the support from the RFBR, grant 19-07-00406.
}


\section*{References}

\begin{thebibliography}{26}

\bibitem{Zhongkai} Zhongkai Huang, Masayuki Hoshina, Hajime Ishihara, and Yang Zhao, Transient dynamics of super Bloch oscillations of a one dimensional Holstein polaron under the influence of an external AC electric field, \textit{Annalen der Physik}, 2017, V. 529, 1600367, DOI:	 10.1002/andp.201600367

\bibitem{D.Hennig} D. Hennig, A. D. Burbanks, and A. H. Osbaldestin, Directed current in the Holstein system, \textit{Phys. Rev. E}, 2011, V. 83, 031121,  DOI:https://doi.org/10.1103/PhysRevE.83.031121

\bibitem{Yakushevich} Yakushevich L.V., Balashova V.N., Zakiryanov F.K., On the DNA Kink Motion Under the Action of Constant Torque, \textit{Math. Biol. Bioinf.}, 2016, V. 11, issue 1, P. 81-90. doi: 10.17537/2016.11.81

\bibitem{STARIKOV_IJMP} E.B. Starikov, J.P. Lewis, O.F. Sankey. Base sequence effects on charge carrier generation in DNA: a theoretical study. \textit{International Journal of Modern Physics B.} 2005. V. 19. issue 29. P. 4331--4357. DOI: 10.1142/S0217979205032802

\bibitem{5}	P.J. De Pablo et. al. Absence of dc-Conductivity in $\lambda$--DNA. \textit{Phys. Rev. Lett.} 2000. V. 85. P. 4992--4995. https://doi.org/10.1103/PhysRevLett.85.4992

\bibitem{6}	D. Porath, A. Bezryadin, S. De Vries, C. Dekker. Direct measurement of electrical transport through DNA molecules. \textit{Nature}. 2000. V. 403. P. 635--638. https://doi.org/10.1038/35001029

\bibitem{Value1} A.A.  Voityuk, N. R\"{o}sch, M. Bixon, J. Jortner. Electronic Coupling for Charge Transfer and Transport in DNA. \textit{J. Phys. Chem. B}. 2000. V. 104. I. 41. P. 9740--9745. https://doi.org/10.1021/jp001109w

\bibitem{Kasumov} A.Y. Kasumov et. al., Proximity-Induced Superconductivity in DNA. \textit{Science}. 2001. V. 291. I. 5502. P. 280--282. doi:10.1126/science.291.5502.280

\bibitem{9}	A. Chepeliaskii et. al. Conduction of DNA molecules attached to a disconnected array of metallic Ga nanoparticles. \textit{New J. Phys.} 2011. V. 13. P. 063046. https://doi.org/10.1088/1367-2630/13/6/063046


\bibitem{10} D. Porath, G. Cuniberti, R. Di Felice. Charge transport in DNA-based devices. \textit{Top. Curr. Chem.} 2004. V. 237. P. 183--227. http://dx.doi.org/10.1007/b94477

\bibitem{Chetverikov} A. P. Chetverikov, W. Ebeling, V. D. Lakhno, and M. G. Velarde, Discrete-breather-assisted charge transport along DNA-like molecular wires. \textit{Phys. Rev. E}, 2019, V. 100, 052203, DOI:https://doi.org/10.1103/PhysRevE.100.052203

\bibitem{11} R.G. Eudres, D.L. Cox, R.R.P. Singh. Colloquium: The quest for high-conductance DNA. \textit{Rev. Mod. Phys.} 2004. V. 76. P. 195--214. http://dx.doi.org/10.1103/RevModPhys.76.195

\bibitem{14} M.Taniguchi, T.Kawai. DNA electronics. \textit{Physica E}. 2006. V. 33. P. 1--12. https://doi.org/10.1016/j.physe.2006.01.005


\bibitem{15} E.M. Conwell, S.V.Rakhmanova. Polarons in DNA. \textit{Proc. Natl. Acad. Sci.} 2000. V. 97. P. 4556--4560.  https://doi.org/10.1073/pnas.050074497

\bibitem{Voulgarakis} Nikolaos K. Voulgarakis. The effect of thermal fluctuations on Holstein polaron dynamics in electric Field.    \textit{Physica B}, 2017 V. 519, P. 15--20  http://dx.doi.org/10.1016/j.physb.2017.04.030

\bibitem{Nazareno} H. N. Nazareno, P.E. de Brito, Bloch oscillations as generators of polarons in a 1D crystal, \textit{Physica B}, 2016, V. 494, P. 1--6. doi:10.1016/j.physb.2016.04.029


\bibitem{Fialko2019} N.S. Fialko, V.D. Lakhno, Dynamics of Large Radius Polaron in a Model Polynucleotide Chain with Random Perturbations. \textit{Math. Biol. Bioinf.}, 2019, V. 14 issue 2, P.406-419, doi: 10.17537/2019.14.406

\bibitem{Fuentes} M. A. Fuentes, P. Maniadis, G. Kalosakas, Rasmussen, A. R. Bishop, V. M. Kenkre, and Yu. B. Gaididei. Multipeaked polarons in soft potentials, \textit{Phys. Rev. E}, 2004, V. 70, 025601(R), DOI:10.1103/PhysRevE.70.025601

\bibitem{Vinogr2019} Astakhova T.Yu., Vinogradov G.A. Polaron in Electric Field and Vibrational Spectrum of Polyacetylene. \textit{Math. Biol. Bioinf.} 2019. V. 14. issue 1. P. 150--159. Published in Russian. doi: 10.17537/2019.14.150

\bibitem{Lak2000} V.D. Lakhno. Soliton-like Solutions and Electron Transfer in DNA. \textit{J. Biol. Phys.} 2000. V. 26. P. 133--147. https://doi.org/10.1023/A:1005275211233

\bibitem{Holstein} T. Holstein. Studies of polaron motion: Part I. The molecular-crystal model. \textit{Annals of Phys}. 1959. V. 8. P. 325--342. doi: 10.1016/0003-4916(59)90002-8

\bibitem{Holstein2} T. Holstein. Studies of polaron motion: Part II. The "small" polaron. \textit{Annals of Phys}. 1959. V. 8. P. 343--389. DOI: 10.1016/0003-4916(59)90003-X


\bibitem{jtf1} A.N. Korshunova, V.D. Lakhno. Simulation of the Stationary and Nonstationary Charge Transfer Conditions in a Uniform Holstein Chain Placed in Constant Electric Field. \textit{Technical Physics}. 2018, Vol. 63. issue 9. P. 1270--1276. \\DOI: 10.1134/S1063784218090086

\bibitem{PhE} A.N. Korshunova, V.D. Lakhno. A new type of localized fast moving electronic excitations in molecular chains. \textit{Physica E}. 2014. V. 60. P. 206--209. \\DOI: 10.1016/j.physe.2014.02.025

\bibitem{LaKo2011} V.D. Lakhno, A.N. Korshunova. Electron motion in a Holstein molecular chain in an electric field. \textit{Euro. Phys. J. B}. 2011. V. 79, P. 147--151. \\DOI: 10.1140/epjb/e2010-10565-2

\bibitem{LaKo2007} V.D. Lakhno, A.N. Korshunova. Bloch oscillations of a soliton in a molecular chain. \textit{Eur. Phys. J. B}.  2007. V. 55. 85--87. https://doi.org/10.1140/epjb/e2007-00045-3


\end{thebibliography}
\end{document}